# Label-free identification of individual bacteria using Fourier transform light scattering

YoungJu Jo,[1] JaeHwang Jung,[1] Min-hyeok Kim,[2] HyunJoo Park,[1] Suk-Jo Kang,[2] and YongKeun Park[1,*]

[1]Department of Physics, Korea Advanced Institute of Science and Technology, Daejeon 305-701, Republic of Korea
[2]Department of Biological Sciences, Korea Advanced Institute of Science and Technology, Daejeon 305-701, Republic of Korea
*yk.park@kaist.ac.kr

**Abstract:** Rapid identification of bacterial species is crucial in medicine and food hygiene. In order to achieve rapid and label-free identification of bacterial species at the single bacterium level, we propose and experimentally demonstrate an optical method based on Fourier transform light scattering (FTLS) measurements and statistical classification. For individual rod-shaped bacteria belonging to four bacterial species (*Listeria monocytogenes*, *Escherichia coli*, *Lactobacillus casei*, and *Bacillus subtilis*), two-dimensional angle-resolved light scattering maps are precisely measured using FTLS technique. The scattering maps are then systematically analyzed, employing statistical classification in order to extract the unique fingerprint patterns for each species, so that a new unidentified bacterium can be identified by a single light scattering measurement. The single-bacterial and label-free nature of our method suggests wide applicability for rapid point-of-care bacterial diagnosis.

## 1. Introduction

Rapid and accurate identification of bacterial species is crucial for diagnosing infectious diseases or screening for poisoning sources in food. Bacterial infection causes a number of severe diseases and syndromes, e.g. sepsis, meningitis, pneumonia, and gastroenteritis, which all require immediate and appropriate treatment based on the detection and identification of the bacterial species [1-3]. However, conventional techniques for bacterial identification have significant limitations, mainly due to the long procedural time required for identification. The present gold standard, bacterial blood culture followed by susceptibility testing for drug resistance, requires days for preliminary results and is often negative for fulminant cases. Recently-developed methods based on real-time quantitative polymerase chain reaction (qPCR) followed by sequencing are faster and more robust; however, they still require hours and are often too expensive for resource-limited environments [4-6]. Consequently, the treatment of many important diseases and syndromes (such as sepsis, of which mortality rates increase by approximately 9% per hour before appropriate antibiotic therapy [7]) typically employs wide-spectrum antibiotic therapy based on clinical experience. This strategy is inefficient compared to accurately species-targeted therapy and carries the potential dangers of severe side effects or the emergence of antibiotic-resistant microorganisms [8, 9].

Various non-invasive optical approaches have been studied to address this issue, as the root cause limiting the speed of the aforementioned biochemical techniques is lengthy and intensive sample processing. Among spectroscopic methods, Fourier transform infrared spectroscopy (FTIR) of bacterial colonies [10-12] and Raman scattering (and also surfaced-enhanced Raman scattering) of individual bacteria [13-16] were extensively investigated. The FTIR approach simplifies the sample processing compared to biochemical methods, though its speed is still limited, as the colonies should be cultured. The

Raman scattering method presents excellent sensitivity of the single bacterium level that enables bypassing the culturing step; however, the signals are too weak for high-throughput investigation. Employing the surface enhancement effect compensates for this shortcoming at the cost of labeling with exogenous agents such as metallic nanostructures [17].

In addition to the spectroscopic methods, angle-resolved light scattering (ALS) based on *spatial* light information has been investigated as well at three different levels: bacterial suspensions, bacterial colonies, and individual bacteria. ALS of bacterial suspensions was pioneered by Wyatt and colleagues in the 1960's [18-20] and has been intensively studied in recent years [21-26]. This ALS approach possess a similar advantage and disadvantage with those of FTIR: simplified but still lengthy sample preparation. ALS of individual bacteria has received relatively little research attention [27-31], probably because certain characteristics of single-bacterial ALS measurement (extremely small scattering cross section, wide scattering angle range, and significantly high dynamic range) make this method technically challenging [32]. Despite these difficulties, ALS of individual bacteria promises a powerful advantage: label-free identification at the single bacterium level.

Here, we present a novel ALS-based identification of individual bacteria using Fourier transform light scattering (FTLS). Quantitative phase imaging (QPI) of individual bacteria in combination with FTLS technique enables single-shot measurement of single-bacterial two-dimensional (2D) ALS maps covering a broad angular range with unprecedented precision, as recently demonstrated by our group [32]. The measured ALS maps, which essentially reflect cellular and subcellular structures and compositions [33], are then analyzed for statistical classification; the ensembles of ALS maps from multiple bacterial species are systematically analyzed to extract the unique fingerprint patterns for each species, so that a new unidentified bacterium can be identified by a single light scattering measurement. Our experimental results demonstrate that detectable and significant patterns in the 2D light scattering spectra can distinguish individual bacteria belonging to four rod-shaped species (*Listeria monocytogenes*, *Escherichia coli*, *Lactobacillus casei*, and *Bacillus subtilis*), which are indistinguishable with only cellular shapes, achieving cross-validation accuracy higher than 94%. The single-bacterial and label-free nature of our method suggest wide applicability for rapid point-of-care bacterial diagnosis.

## 2. Methods and results

*2.1. Model problem: Four rod-shaped bacterial species*
In order to demonstrate the capability of the present approach, we set and solve a virtual clinical case for the proof-of-concept study: the objective is to identify the species of an unidentified bacterial pathogen, i.e. a single specific bacterium isolated from the sample that is known to belong to one of the four rod-shaped bacterial species, *Listeria monocytogenes. Escherichia coli, Lactobacillus casei, and Bacillus subtilis.* These four species, including both Gram-positive and Gram-negative ones, were chosen for this study as the rod-shaped species are clinically important in general, while they exhibit similar shapes and sizes that make them indistinguishable via simple observation of cellular morphology using conventional bright-field microscopy or phase contrast microcopy. Although this model problem requires *a priori* knowledge that the target bacterium belongs to one of the four species, the method described throughout this report can be readily extended to a more general approach (see Discussion and Conclusions).

In order to demonstrate that the simple morphological information is not sufficient to identify the bacterial species, we illustrate the size distribution of the individual bacteria (characterized by the lengths of major and minor axes of the rod-shaped bacterial cells) in unsynchronized growth states belonging to the four species in Fig. 1 (see the following sections for the imaging method). The size and aspect ratio of the cells in these species are similar to one another, and this result clearly implies that it is impractical to distinguish these bacterial species using morphological parameters, which is the typical accessible information using conventional bright-field microscopy or phase contrast microscopy.

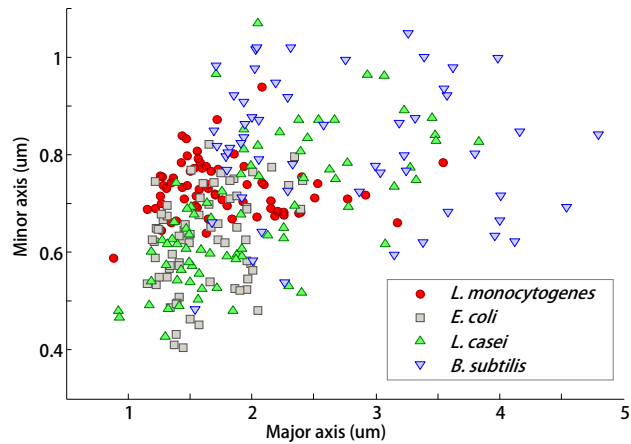

Fig. 1. The size distribution of the four bacterial species, *L. monocytogenes*, *E.coli*, *L. casei*, and *B. subtilis* (67, 69, 78, and 55 bacterial cells, respectively). The lengths of the major and minor axes of each rod-shaped bacterium [indicated in Fig. 2(a)] in unsynchronized growth states are observed and plotted (see the following sections for the imaging method). It is impractical to identify the species from only cellular shapes, which are observable using conventional optical microscopy.

*2.2. Preparation of individual bacteria samples*
Four bacterial species, *L. monocytogenes*, *E. coli*, *L. casei*, and *B. subtilis*, were prepared according to the following procedures.
- *L. monocytogenes* strain (10403S) was grown on Brain-Heart Infusion agar plates without antibiotics.
- *E. coli* strain (ER 2738) with tetracycline-resistance was grown on Luria-Bertani agar plates with tetracycline (50 mg/mL).
- *L. casei* strain (KCTC 2180) was grown on MRS agar plates without antibiotics.
- *B. subtilis* strain (KCTC 1023) was grown on nutrient agar plates without antibiotics.

After overnight culturing in a 37°C incubator, a few tips of solid-cultured bacterial colonies were taken and then put into the respective Eppendorf-tubes. A Dulbecco's Phosphate Buffered Saline solution (LB 001-12, Welgene, Republic of Korea) was added in each tube to prepare a bacterial solution. After slight vortex mixing, a small volume (10 μL) of the bacterial solution was sandwiched between standard microscopic cover glasses (C024501, Matsunami Glass Ind., Japan) with a spacer made of double-sided tape with a thickness of 20-30 μm. Imaging was performed after the bacterial cells settled down to the bottom, but no later than 15-20 minutes after preparation. The aforementioned dilution step was performed until the bacterial cells were spread into a single layer when being imaged.

*2.3. Quantitative phase imaging of individual bacteria*
In order to precisely measure the 2D light scattering maps from individual bacteria, we employed QPI and FTLS techniques [32, 34-37]. For the first step, the optical field maps (both amplitude and phase maps of the transmitting light field) of isolated individual bacteria are precisely measured using QPI [34, 35]:

$$E(\vec{r}) = A(\vec{r})\exp\left[i\Delta\phi(\vec{r})\right]$$,

where $A(\vec{r})$ and $\Delta\phi(\vec{r})$ are the amplitude and phase delay maps at position $\vec{r}$, respectively. The phase delay $\Delta\phi(\vec{r})$ results from the relative optical path length, the line integral of the refractive index (RI) through the sample, at each lateral position. At room temperature, we imaged 67, 69, 78, and 55 individual bacterial cells for *L. monocytogenes*, *E.coli*, *L. casei*, and *B. subtilis*, respectively, which were prepared as described in the previous section as unsynchronized growth states.

Here we employed diffraction phase microscopy (DPM), one of the QPI techniques, for optical field imaging [38, 39]. DPM is a common-path interferometry with off-axis spatial modulation, and it provides single-shot full-field imaging capability and high sensitivity for phase measurement. Briefly, a diode-pumped solid state laser ($\lambda$ = 532.1 nm, Cobolt Samba, Sweden) was used as the illumination source. An inverted microscope (IX71, Olympus American Inc., USA), equipped with a high numerical aperture (NA) objective lens (UPLFLN 60X, 1.42 NA, oil-immersion, Olympus American Inc.), was modified with additional optics to be used for a DPM setup. With the additional relay optics, the overall magnification of the system was ×200. A CMOS camera (Neo sCMOS, Andor, UK) was used to record the holograms of the samples. The optical field images were reconstructed from the recorded holograms using a custom-built field retrieval algorithm implemented in MatLab [40, 41]. Each bacterial cell in a field-of-view can be isolated by generating image windows based on phase values for subsequent single-bacterial analyses. More details on the setup and principles of DPM can be found elsewhere [38, 39].

The representative amplitude $A(\vec{r})$ and phase $\Delta\phi(\vec{r})$ images of an individual *L. moncytogenes* bacterium are illustrated in Figs. 2(a) and 2(b), respectively. The images of bacteria in *E. coli*, *L. casei*, and *B. subtilis* are compatible with the case of *L. monocytogenes* [32]. The importance of simultaneous measurements of both maps are described in the following sections.

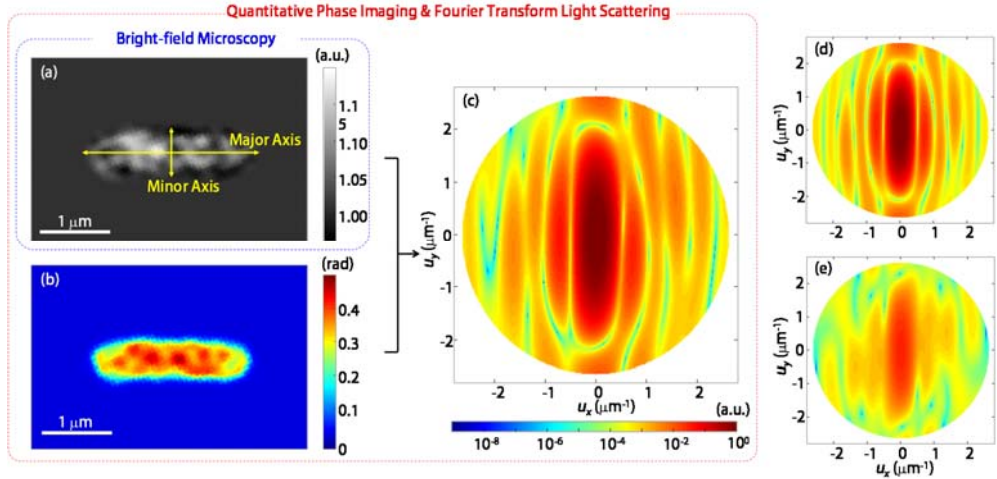

Fig. 2. QPI and FTLS of individual bacterium. (a) Amplitude and (b) phase images of an isolated *L. monocytogenes* bacterium observed using QPI. (c) The corresponding 2D light scattering map generated by FTLS using the full optical field information from (a) and (b). The pseudo-FTLS maps by numerical propagation (d) without amplitude information, and (e) without phase information. The results indicate that phase information, a unique feature of QPI, is central to single-bacterial characterization rather than easily accessible amplitude information.

*2.4. Fourier transform light scattering of individual bacteria*

The measured optical field maps of individual bacteria $E(\vec{r})$ are utilized to calculate the single-bacterial ALS maps using FTLS technique [36, 37]. FTLS is based on the numerical propagation of the optical field in the sample plane by simply Fourier transforming it to generate the far-field light scattering map:

$$I(\vec{u}) = \frac{1}{2\pi}\left|\iint E(\vec{r})\exp(-2\pi i\vec{u}\cdot\vec{r})d^2\vec{r}\right|^2,$$

where $\vec{u}$ is the lateral spatial frequency vector.

A single QPI measurement enables us to attain the corresponding 2D light scattering map at a broad angular range [36, 37]. FTLS is the *spatial* analogy of FTIR, so it provides an unprecedented signal-to-noise ratio due to Fellgett's advantage [42]. The upper limit of detectable light scattering angle is determined by the NA of the objective lens; our system effectively covers the scattering angle ranging from -70° to 70° [32]. It is also worth noting that the measured optical field maps of isolated individual bacteria are numerically centered, rotated, and zero-padded prior to Fourier transformation; this process ensures species identification based on light-cell interaction, independent of the orientation of cells. The scale of zero padding should be appropriately chosen for sufficient but not excessive (a large number of pixels is computationally expensive for the following image analyses) angular resolution of the FTLS maps.

Figure 2(c) illustrates the measured light scattering map of the isolated individual bacterium presented in Figs. 2(a)–(b). The unique FTLS map of each bacterium essentially results from its cellular and subcellular structures and compositions because the light scattering behavior is determined by the 3-D spatial distribution of RI of the cell. The high sensitivity of RI to the biochemical compositions and their local concentrations is central to the effectiveness of FTLS for the study of pathophysiology of cells and tissues [34-37, 43-45].

To re-emphasize that the FTLS maps are not merely determined by the cellular morphology, which can also be obtained with conventional bright-field microscopy or phase contrast microscopy, we calculated the pseudo-FTLS maps of the individual bacterium, as shown in Figs. 2(d)–(e). The numerical light field propagations were performed without amplitude information, $A(\vec{r})=1$ and without phase information $\Delta\phi(\vec{r})=0$, as illustrated in Figs. 2(d) and 2(e), respectively. This result clearly demonstrates that the optical phase information, which is the unique advantage of QPI, is crucial for obtaining the correct ALS spectra. and the amplitude only information only provide limited information.

*2.5. Statistical classification of FTLS maps: Overall procedure*

The measured FTLS maps were then systematically analyzed in order to extract the unique fingerprint patterns for each bacterial species so that a new unidentified bacterium can be identified by a single FTLS measurement. This training-and-identification strategy is called statistical classification, or supervised machine learning, which is one of the most rapidly evolving fields in computer science with diverse applications including bioinformatics [46]. Statistical classification has been previously exploited in various forms in most of the optical methods for bacterial identification briefly reviewed in the Introduction [10-16, 21-23, 26, 28-31].

The overall procedure for the proposed identification scheme is summarized in Fig. 3. After the FTLS measurement, as described in previous sections, the training step is performed as follows. First, the variables describing each FTLS map, which are called features, are extracted by principal component analysis (PCA). The extracted features are then selected and optimized to construct a statistical classification model (classifier) based on linear discriminant analysis (LDA). This PCA-LDA method has been widely utilized in several statistical classification problems [47]. Here, this PCA-LDA process is conducted for species classification; the unique fingerprint patterns from light scattering spectra are efficiently extracted by selecting the species-dependent features and excluding the features with high relevance to the cell-to-cell shape variations within each species that arise from cell growth and division. After completion of the training step, a new unidentified bacterium can be instantly identified by a single FTLS measurement. We assess the identification accuracy via cross-validation. The following sections describe each step of this process in detail.

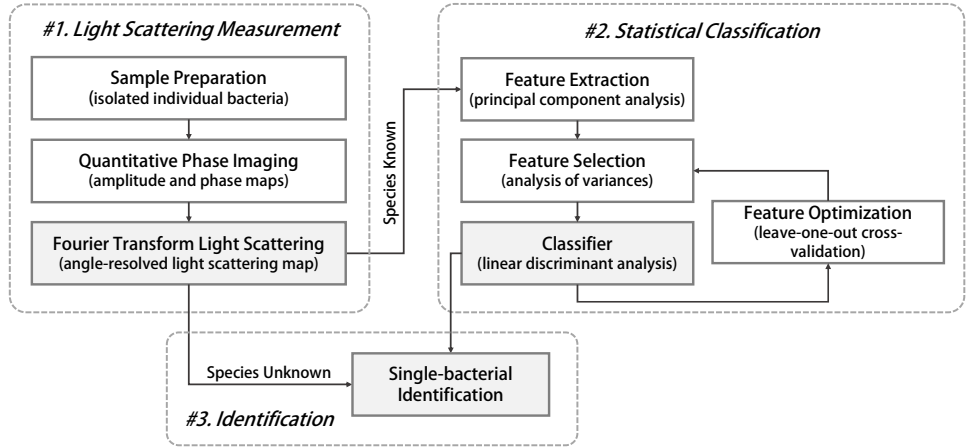

Fig. 3. Overall procedure for single-bacterial identification. 2D ALS maps of isolated individual bacteria are measured using FTLS. The measured FTLS maps are then systematically analyzed in order to extract the unique fingerprint patterns for each species (classifiers) so that a new unidentified bacterium can be identified by a single light scattering measurement.

*2.6. Feature extraction from light scattering spectra*

In order to effectively extract the features from light scattering spectra, the region of interest (ROI) containing the bandwidth limited information (i.e. scattered light collected by the objective lens with specific NA) in the measured 2-D FTLS maps is selected. As illustrated in Fig. 2(c), the ROI is defined as a circle whose diameter corresponds to the maximum spatial frequency, which corresponds to $NA/\lambda$ where $\lambda$ is the wavelength of the illumination. For the QPI system used in this study, a typical ROI is composed of approximately $10^4$ pixels. This high dimensionality of the ROI is computationally expensive, and the analysis would suffer from the 'curse of dimensionality', meaning that the predictive power of classifiers decreases as the dimensionality increases [48]. Thus, the number of variables describing the ROI must be adequately reduced in the following feature extraction procedure.

In order to represent a FTLS map as a linear combination of a small but sufficient number of uncorrelated principal patterns, we employed PCA [49]. In principle, the number of principal patterns used to express the scattering spectra is the same as the pixel numbers in the ROI, because PCA is essentially an orthogonal basis transformation. However, because most information or variance of the original pattern is contained in the first few principal patterns after the PCA, we can significantly reduce the dimensionality to manageable numbers for statistical classification. For this study, we performed PCA on the single-bacterial FTLS maps composed of 5,525 pixels each to generate 250 principal patterns and the corresponding principal pattern coefficients, preserving 99.9% of the original variance or information. The first 16 principal patterns are presented in Fig. 4(a), and the full spectra of the principal pattern coefficients for all investigated individual bacteria are illustrated in Fig. 4(b). Well-established mathematical details of PCA and its applications for biological image analysis can be found elsewhere [49-51].

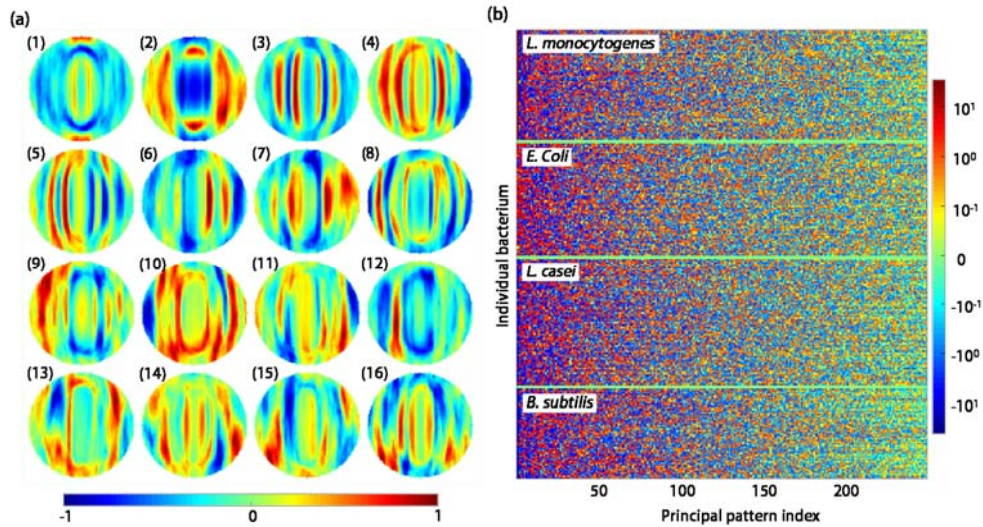

Fig. 4. PCA of the FTLS patterns to extract the features for training the classifiers. (a) The first 16 principal patterns generated by PCA of all measured FTLS patterns. Each principal pattern is scaled into [−1, 1] for visualization purposes. (b) The full spectra of the principal pattern coefficients up to the 250th coefficient. Each row represents an individual bacterium, whereas each column represents a principal pattern. The principal patterns included here contain 99.9% of the original variance or information.

*2.7. Training classifiers with optimized features*

Our final goal is to construct the classification model (classifier) using the extracted features from the FTLS maps. Here we employed LDA, one of the most traditional and robust linear classifiers [52, 53]. LDA draws linear discriminant hyperplanes as optimal boundaries between the bacterial species when each individual bacteria is expressed as a point in the feature space. The features exploited for LDA should be carefully selected and optimized to maximize the accuracy of identification as described below. Since LDA is a linear classifier, the computation is sufficiently fast to enable the iterative construction of the classification models with various subsets of features and samples.

After constructing a classification model, we can assess the accuracy of single-bacterial identification via leave-one-out cross-validation [54]. We construct the classifier using the data from all bacterial cells except one and then apply the classifier to check if it identifies the excluded bacterium correctly. Through repeating this process for all individual bacteria, we can measure the identification accuracy. It is mathematically proven that leave-one-out cross-validation is a precise estimation of the real accuracy to identify independently measured data [46].

For the first step for feature selection, we rank the principal patterns according to their usefulness for statistical classification. Because principal patterns are linearly uncorrelated, we can independently investigate each principal pattern, as shown in each column of Fig. 4(b). We employed the analysis of variances (ANOVA), which is a generalization of the Student's *t*-test into more than two groups, to measure the species-distinguishing capability of each principal pattern [55]. Since the values in each column of Fig. 4(b) belong to one of the four species, we can calculate the *p*-value between the species for each column that estimates the discriminating power of each principal pattern. In Fig. 5(a), we sorted the principal patterns in the order of ascending *p*-values. The principal patterns with *p*-values close to zero are more informative for classification, whereas the principal patterns with *p*-values close to one are noisy (i.e. these patterns may result in overtraining that lowers accuracy [46]), as illustrated in Figs. 5(b) and (c), respectively. This result establishes a heuristic strategy for feature selection [46]: if the principal patterns are added in a consecutive or accumulative manner, from a low principal pattern index to a high principal pattern index (sorted by *p*-values), as features for statistical classification, the accuracy of the identification based on each classification model will be maximized at a certain optimal point in the middle. This expected tendency was observed in our data, as shown in Fig. 5(d).

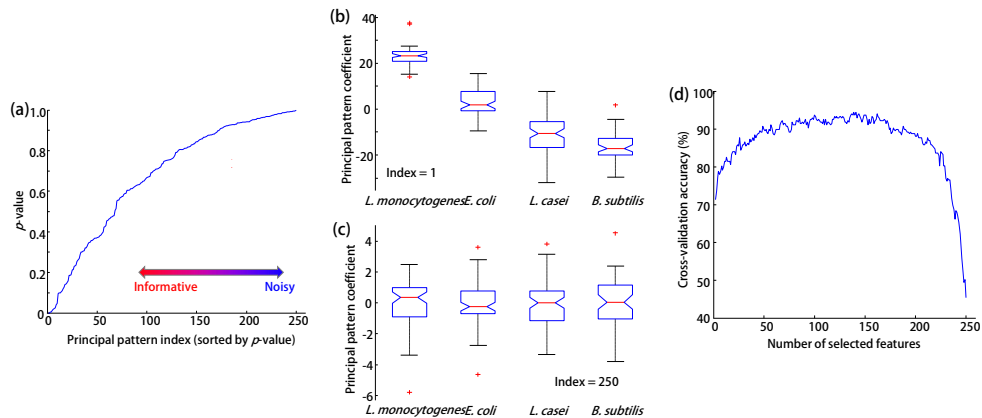

Fig. 5. Feature selection and optimization for training the classifiers. (a) The *p*-value of each principal pattern coefficient as a measure of species-distinguishing capability is calculated using ANOVA and then the principal patterns are sorted in the order of ascending *p*-values. The principal pattern coefficients are presented in box plots for (b) the most distinguishable (index = 1) and (c) the most indistinguishable (index = 255) principal patterns. The first principal patterns with *p*-values close to zero are informative for classification, while the last principal patterns with *p*-values close to one are noisy. Thus, we select the principal patterns in a consecutive or accumulative manner, from left to right, as the features for statistical classification. The cross-validation accuracy for the classifiers trained in this manner is plotted in (d), where the optimal classification is achieved in the middle as expected.

The coarse-grained cross-validation accuracy consists of eight detailed parameters (two parameters per species) of identification accuracy: sensitivity (true positive results over all positive inputs) and specificity (true negative results over all negative inputs) for each species. Selecting a classification model with high values of some parameters may give low values for other parameters, and vice versa. Thus, the final selection of the classifier (or feature optimization for classifier training) should consider the specific purpose of application. For example, we might be interested simply in whether a given bacterium is pathogenic *L. monocytogenes* or not (see Discussions and Conclusions).

The identification results are illustrated in Fig. 6. Here, we selected a classification model with homogeneously high values of the accuracy parameters. The overall cross-validation accuracy was 94.05%, with sensitivities of 95.52%, 95.65%, 88.46%, and 98.18% and specificities of 99.51%, 96.50%, 97.91%, and 98.13% for *L. monocytogenes*, *E. coli*, *L. casei*, and *B. subtilis*, respectively.

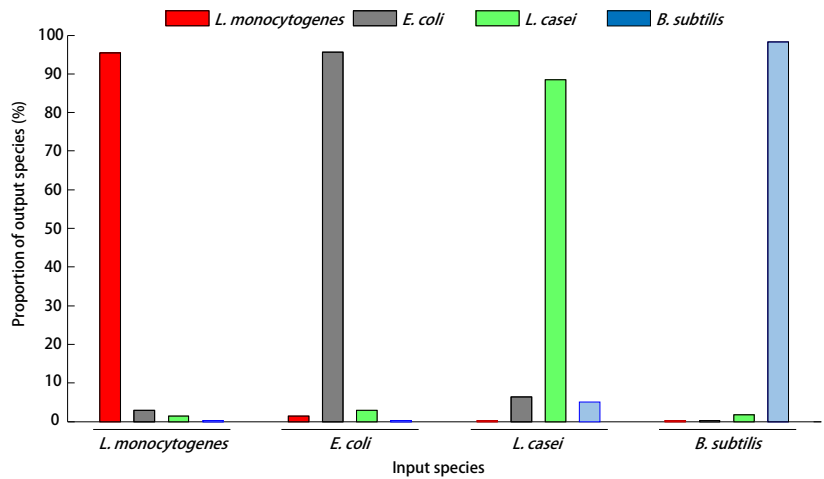

Fig. 6. Single-bacterial identification with the optimized classifier. The proportion of each output species for a certain input species, where leave-one-out cross-validation is utilized to precisely mimic the independently measured data, is plotted. The overall accuracy was 94.05%, with sensitivities of 95.52%, 95.65%, 88.46%, and 98.18% and specificities of 99.51%, 96.50%, 97.91%, and 98.13% for *L. monocytogenes*, *E. coli*, *L. casei*, and *B. subtilis*, respectively.

## 3. Discussions and conclusions

We have presented a novel optical method for label-free species identification of individual bacteria using FTLS and statistical classification. Four rod-shaped bacterial species (*L. monocytogenes*, *E. coli*, *L. casei*, and *B. subtilis*) were indistinguishable using only their cellular shapes due to the limited information in intensity images and the cell-to-cell variations arising from cell growth and division. To address this difficulty in species classification, we have exploited the single-bacterial 2D ALS maps measured by FTLS to systematically extract the unique fingerprint patterns for each species, or the optimized LDA classifier. Then, a new unidentified bacterium can be identified by a single light scattering measurement followed by application of the classifier with accuracy higher than 94% as assessed by cross-validation.

The single-bacterial and label-free nature of our method implies wide applicability for rapid point-of-care bacterial diagnosis. As briefly reviewed in the Introduction, the ALS of individual bacteria has significant advantages for rapid and label-free identification of bacteria, as compared to other existing optical or biochemical approaches. Though several groups have demonstrated similar identification schemes based on single-bacterial ALS measurement and statistical classification [28-31], the present approach provides ALS measurements in a wide angular range with unprecedented precision [32]. The importance of broad angular range (especially high scattering angle) for bacterial species characterization lies in the capability of accessing the light interacting with small subcellular organelles, which is analogous to side light scattering in flow cytometry analysis. In addition, the present method is based on a single-shot measurement; after the statistical classification for the target species, only a single-shot FTLS measurement per bacterium is required for species identification. Simply attaching a compact QPI unit to a conventional high-speed imaging flow cytometry will enable high-throughput analysis for immediate and efficient therapy [56]. This system can be even miniaturized for point-of-care diagnosis [57].

In particular, the present single bacterium approach can be exploited for identifying unculturable bacteria and avoiding inter-species competition while culturing heterogeneous populations. Furthermore, our concepts can be readily extended to other morphologies such as spherical and spiral bacterial species and will hopefully lead to the addition of other types of microbes or general pathogens to the optical identification regime. Moreover, the non-invasive nature of the proposed method enables parallel investigations for more detailed single bacterium profiling. For instance, the use of various QPI modalities, such as spectral [58-60], polarimetric [61], synthetic [62], tomographic [63-65], and reflection [66] QPI, will further improve the accuracy and applicability of the proposed method. For more efficient therapy, our method might be co-administered with more robust genotype-oriented qPCR methods [67] or species-independent bacterial filtering therapy [68].

There are several limitations of the present method to be addressed in future works. As is common for all single cell approaches, the purification of samples is required to avoid interference by non-bacterial or abiotic components. We expect that droplet microfluidics on a chip might be a solution compatible with the aforementioned point-of-care scheme [69]. In addition, while differentiating viable bacterial cells from dead ones is important for clinical setting, both optical and biochemical methods have limitations.

In addition to the technical issues, there are several points to be improved for constructing more a robust bacterial ALS fingerprint database for the practical implementation of our method. Here, we demonstrated the systematic extraction of species-dependent information through overcoming the cell-to-cell shape variations, which are 'noise' to identification. However, the following noises also exist and must also be overcome: cell strains within species and environmental factors that alter phenotypes or gene expression including temperature [70], pH [71], osmolality [72], and host-pathogen interactions [73]. Experiments with diverse cell strains under various physiological conditions should be performed in order to suppress these noises and to construct a robust fingerprint database prior to practical implementation of the present method. Since many noise sources contribute simultaneously, employing nonlinear classifiers, such as artificial neural networks [74] or nonlinear support vector machines [75], can be employed to improve the accuracy of identification. Furthermore, the current requirement of *a priori* knowledge of the species range also needs to be overcome. One strategy is to focus on several pathogenic species that cause major diseases. Another strategy is tree-structured phenotype classification: constructing a classifier for bacteria or non-bacteria, a classifier for morphological categories, a classifier for Gram-positive or Gram-negative bacteria, a classifier for genus, and so on. Tree-structured classification strategy is known to be effective for multiclass classification problems [47]. All of these routine tasks can be conducted in a specialized lab and can be uploaded on the web for wide usage of light scattering spectra for bacterial identification.

**Acknowledgments**
The authors thank Ms. Minryung Song (Department of Bio and Brain Engineering, KAIST) for her kind advice on machine learning algorithms. This work was supported by KAIST Undergraduate Research Participation (URP) program, the Korean Ministry of Education, Science and Technology (MEST), and the National Research Foundation (2012R1A1A1009082, 2013M3C1A3063046, 2012-M3C1A1-048860, 2013K1A3A1A09076135, NRF-2012M3A9B4027955, NRF-2011-0020334, NRF-2010-0009042). Y.J. acknowledges support from KAIST Presidential Fellowship and SPIE Optics & Photonics Education Scholarship.